%% file: amtc_arxiv.tex
\documentclass[journal,10pt]{IEEEtran}

\usepackage{graphicx,xcolor}
\usepackage{nicefrac,xfrac}
\usepackage{lipsum}
\usepackage{amsmath,amssymb,amsfonts,accents,mathrsfs, nccmath}

\def\bxi{{\boldsymbol{\xi}}}
\def\bphi{{\boldsymbol{\phi}}}
\def\bPhi{{\boldsymbol{\Phi}}}
\def\bH{{\mathbf{H}}}
\def\bh{{\mathbf{h}}}
\def\bY{{\mathbf{Y}}}

\def\bS{{\mathbf{S}}}

\begin{document}
%
\title{\LARGE Dynamic Message Scheduling With  Activity-Aware Residual Belief Propagation for Asynchronous mMTC Systems}

\author{Roberto~B.~Di~Renna,~\IEEEmembership{Student~Member,~IEEE,}
        and~Rodrigo~C.~de~Lamare,~\IEEEmembership{Senior~Member,~IEEE \vspace{-1.5em}}
\thanks{The authors are with the Centre for Telecommunications Studies (CETUC),
Pontifical Catholic University of Rio de Janeiro (PUC-Rio), Rio de Janeiro 22453-900,
Brazil (e-mail: \{robertobrauer, delamare\}@cetuc.puc-rio.br).
This work was supported by the Conselho Nacional de Desenvolvimento Cient\'{i}fico e Tecnol\'{o}gico (CNPq).}
\thanks{Manuscript received Month XX, 202X; revised Month XX, 202X.}}

\markboth{IEEE Wireless Communications Letters,~Vol.~XX, No.~X, Month~202X}%
{Di Renna \MakeLowercase{\textit{et al.}}: Dynamic Message Scheduling Based on Activity-Aware Residual Belief Propagation for Asynchronous mMTC}

\maketitle

\begin{abstract}
    In this letter, we propose a joint active device detection and channel estimation framework based on factor graphs for asynchronous uplink grant-free massive multiple-antenna systems. We then develop the message-scheduling GAMP (MSGAMP) algorithm to perform joint active device detection and channel estimation. In MSGAMP we apply scheduling techniques based on the residual belief propagation (RBP) and the activity user detection (AUD) in which messages are generated using the latest available information. MSGAMP-type schemes show a good performance in terms of activity error rate and normalized mean squared error, requiring a smaller number of iterations for convergence and lower complexity than state-of-the-art techniques. %
\end{abstract}

\begin{IEEEkeywords}
    mMTC, message-passing, channel estimation, message scheduling, grant-free massive MIMO.
\end{IEEEkeywords}

\IEEEpeerreviewmaketitle

\section{Introduction}
\label{sec:intro}
    \IEEEPARstart{C}overing different industries as healthcare, logistics, process automation and utilities, it is believed that machine-type communications (MTC) will correspond to half of the global connected devices by 2023. Concentrated on the uplink, MTC traffic is typified by small packets transmitted sporadically, often with low data-rate and loose delay constraints~\cite{DiRennaAccess2020}. With these characteristics and the expected huge number of machine-type devices (MTDs), conventional scheduling-based orthogonal multiple access schemes are not suitable.

    A solution  proposed in recent years is based on grant-free \textit{non-orthogonal multiple access} (NOMA)~\cite{MShirvanimoghaddam2017}, where active devices transmit frames without previous scheduling, in order to eliminate the need for round-trip signaling. With the massive number of MTDs requiring access without coordination, a time-slotted transmission would cause significant overhead. In a time-slotted transmission scenario, where devices can change their activity state only at the beginning of each time-slot, any device that fails to align its time slots properly may disturb the whole detection and estimation process. In this way, the study of a non-time-slotted or asynchronous transmission is { promising for mMTC since it has advantages such as reduced transmission latency, smaller signalling overhead due to the simplification of the scheduling procedure and improved energy efficiency (battery life) of MTCDs with the reduction in signalling~\cite{F5G1,CBockelmannAccess2018}. As all MTCDs simply transmit, the work of the BS is increased \cite{mwc,baplnc}, this scenario renders the activity and data detection \cite{spa,mfsic,mbdf} and channel estimation \cite{1bitce} even more challenging tasks.}

    Despite the focus of many works on the joint user activity and data detection problem~\cite{Zhu2011, CWei2017, DiRennaWCL2019,DiRennaTCom2020}, most of these studies considered that the uplink channel state information (CSI) is perfectly known to the base station (BS). However, in practice, the uplink CSI should be estimated before data detection.  Exploiting the \textit{a priori} distribution of the channel sparse vector to be recovered, the works in~\cite{LLiuTSigPr2018,ZChenTWirCom2019,KSenel2018} use compressed sensing (CS)-based techniques in order to assess the channel estimation and the activity error rate (AER) performance. As an extension of the generalized approximate message passing (GAMP) algorithm ~\cite{SRangan2011}, the hybrid GAMP (HyGAMP)~\cite{Rangan2017} exploits the sparsity in the exchange of messages. HyGAMP outperforms other existing algorithms in terms of mean square error (MSE), since it combines a loopy belief propagation (LBP) part for user activity detection and a GAMP-type strategy for channel estimation. However, HyGAMP considers a completely parallel update of the messages, where each iteration performs exactly one update of all edges.

    {In this work, we present a joint active device detection and channel estimation framework based on factor graphs for asynchronous uplink grant-free massive multiple-antenna systems. We also devise the message-scheduling GAMP (MSGAMP) algorithm that uses the factor graph approach and aims to find the best sequence of message updates, improving the convergence and error rates by focusing on the part of the graph that has not converged. Unlike dynamic scheduling techniques \cite{CasadoTCom2010,kaids} used for decoding Low-Density Parity-Check (LDPC) decoders, MSGAMP is applied to factor graphs and performs novel sequential scheduling schemes for message updates in mMTC. In particular, MSGAMP updates messages according to the activity user detection (AUD) and the residual belief propagation (RBP).} Since only a few very recent works~\cite{XMaAccess2020, JZhangOJVT2020,TDingTWC2019} have studied the asynchronous mMTC scenario, MSGAMP addresses the problem of joint active device detection and channel estimation without requiring frame-level synchronization. MSGAMP exploits the \textit{a priori} distribution of the sparse channel matrix and use the number of antennas in the BS to improve the activity detection. Simulations show that MSGAMP results in an improved performance over HyGAMP in terms of normalized MSE (NMSE) with fast convergence and a lower computational cost than existing techniques.

    This paper is structured as follows. Section II introduces the asynchronous system model. The problem of joint channel and user activity estimation along with the proposed MSGAMP is detailed in Section III. Section IV presents the results of simulations, whereas the conclusions are drawn in Section V.

\section{System Model}

    In this section we describe the considered asynchronous grant-free uplink NOMA scenario, where symbol-level synchronization is assumed but not frame-level synchronization. In the uplink, we have $N$ single-antenna MTDs communicating with a BS equipped with $M$ antennas \cite{mmimo,wence}. In the grant-free system model, each frame consists of pilot and data parts~\cite{DiRennaAccess2020}. Since the goal of this work is to jointly detect the activity of devices and estimate their channels, we only consider the part of the frame with pilots. However, the use of the system model for data detection is straightforward.

\begin{figure}[t]
            \centering
            \includegraphics[scale=1]{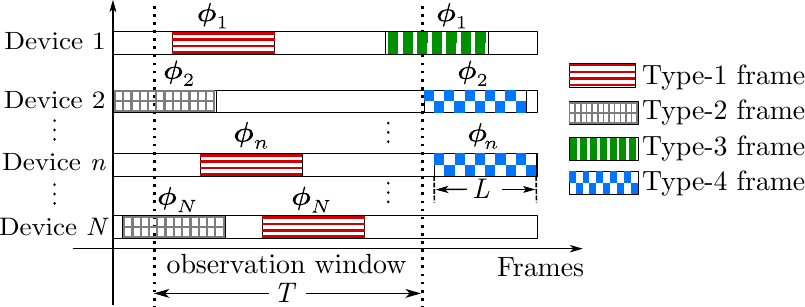}
            \vspace{-5pt}
            \caption{Asynchronous frames of the grant-free random access mMTC scenario.}
            \label{fig:asy_mmtc}
            \vspace{-10pt}
        \end{figure}

    {
    As depicted in Fig.~\ref{fig:asy_mmtc}, at the beginning of any symbol interval, each device is allowed to transmit $L$ pilot symbols, which form a frame. Since mMTC results in sparse systems, we designate the Boolean variable $\xi_{n,t} = 1$ that indicates the activity of the $n$-th device in the $t$-th symbol interval and $\xi_{n,t} = 0$, otherwise. Thus, considering $\rho_n$ the probability of being active of the $n$-th device, $P\left(\xi_{n,t} = 1\right) = 1 - P\left(\xi_{n,t} = 0\right) = \rho_n,$ where all $\xi_{n,t}$ are considered i.i.d. in relation to $n$ and each device has its own activity probability. When an MTD is active, it transmits one of the independent pilot sequences previously provided by the BS. The frame of the $n$-th device is composed by $\bphi_n = \exp{\left(j \pi \boldsymbol{\alpha}\right)}$, where each element of vector $\boldsymbol{\alpha} \in \mathbb{R}^{L}$ is drawn uniformly at random in $\left[-1,1\right]$. Despite the intermittent pattern of transmissions, each device should wait, at least, to the guard period interval to transmit again.

    Let $\bh_t \in \mathbb{C}^{N \times 1}$ be the vector that models the channels between the BS and $N$ devices in the $t$-th symbol interval. Considering $t_n$ as the symbol interval in which the $n$-th device initiates its transmission, each component is modeled as
    \begin{equation}\label{eq:H}
    \resizebox{.9\hsize}{!}{$
        \hspace{-2.6pt}h_{n,t} = \left\{
            \begin{array}{rl}
                \hspace{-5pt}\sqrt{\beta_n}\,a_{n,t}\left(t-t_n+1\right), &\hspace{-2.5pt}\forall\, \left(t_{n} \leq t < t_{n} + L\right), \\
                0, &\hspace{-2.5pt} \text{otherwise.}
            \end{array}\right.$}
    \end{equation}

    \noindent where $\bh_t$ gathers independent fast fading, geometric attenuation and log-normal shadow fading. The vector $\mathbf{a}_{t}$ contains the fading coefficients modeled as circularly symmetric complex Gaussian random variables with zero mean and unit variance, while $\beta_n$ represents the path-loss and shadowing component of each device, which depends on the location of the devices and remains the same for all frames of the $n$-th device.


    As depicted in Fig.~\ref{fig:asy_mmtc}, in the asynchronous scenario, it is possible that just part of the transmitted frame falls within the observation window. Since the problem of interest here is to jointly estimate the channels and the activity of devices, the BS is only able to deal with the type-1 frames. Thus, type-2 and type-3 frames should have their channels estimated and activity detected in another observation window. Accordingly, the BS generates a sequence of observation windows $\left\{t_x, t_x + T\right\}_{x \in \mathbb{Z}_+}$ where $t_x =0$, if $x=1$ and $t_x = t_{x-1} + \Delta t$, otherwise. This sequence can be seen as a sliding window with window size $T$ and step size $\Delta t$. Since $T > L$, any consecutive observation windows have an intersection of $T - \Delta t$ symbol intervals, enabling BS to estimate the channels of all frames.

    Considering the $M$ BS antennas, for an arbitrary observation window $\left[t_x, t_x+T\right)$ and omitting the subscript $t_x$ to simplify the notation, the received signals are described by the model
    \begin{align}\label{eq:sig_model}
        \mathbf{Y}_m &= \boldsymbol{\Phi}\, \mathbf{H}_m + \mathbf{W}_m, \hspace{5pt} \forall \left(m=1,\cdots , M\right)
    \end{align}

    \noindent where $\mathbf{W}_m \in \mathbb{C}^{L\times T}$ is the independent complex-Gaussian noise matrix with $\mathcal{C}\mathcal{N}\left(0,\sigma_w^2\right)$, $\bY_m \in \mathbb{C}^{L\times T}$ is the matrix that gathers the received signals and  $\mathbf{H}_m \in \mathbb{C}^{N\times T}$ the channels. The subscript $m$ indicates which BS antenna received the signals. For each new window, the values of $\mathbf{W}_m$, $\mathbf{Y}_m$ and $\mathbf{H}_m$ change, while $\boldsymbol{\Phi} \in \mathbb{C}^{L\times N}$ keeps the pilot sequence of each device. As in this scenario we have a massive number of devices, the size of the window $T$ is smaller than $N$ thus, the system is overloaded. However, as seen in~(\ref{eq:H}), $\mathbf{H}$ is sparse, which makes its recovery possible through the theory of compressed sensing (CS)~\cite{JWChoi2017}.
} 

\section{Activity Detection and Channel Estimation}
\label{sec:jointAUDCD}
\begin{figure}[t]
            \vspace{-3mm}
            \centering
            \includegraphics[scale=1]{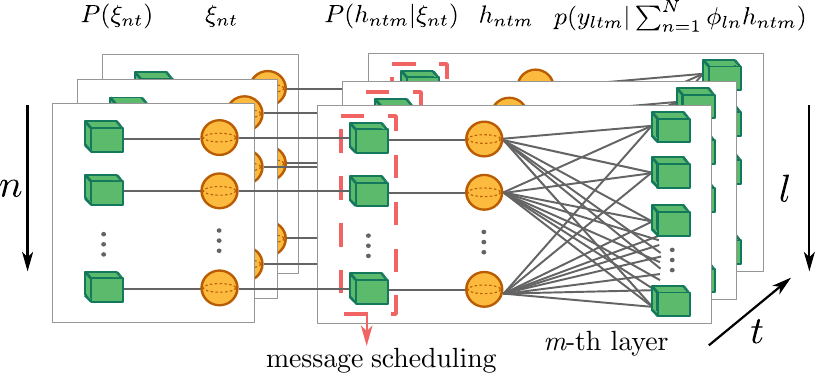}
            \vspace{-2.5mm}
            \caption{The factor graph of joint distribution $p(\mathbf{H},\mathbf{Y},\boldsymbol{\xi})$ where cubes denote factor nodes and spheres variable nodes.}
            \label{fig:fg}%
            \vspace{-5mm}
        \end{figure}

{
       In order to present the message updating rules of the MSGAMP algorithm, we introduce some statistical properties of the system model. Assuming a BS with one antenna ($M=1$), the subscript $m$ is omitted in the following formulation.

    \subsection{Factor Graph approach}

    As reported in the literature~\cite{Rangan2017,QZou2020}, it is possible to estimate the channels exploiting the statistical properties of the system model approximating the marginal posterior density by a product of the prior distribution of $\bh_t$, $p\left(\mathbf{h}_t|\bxi_t\right)$, and the likelihood, $p\left(\bY|\bH,\bxi\right)$. Thus, the minimum MSE (MMSE) estimate of $h_{nt}$, $\hat{h}_{nt} = \mathbb{E}_{h_{nt}|\mathbf{y}}\left[h_{nt}\right] \forall\, n, t$ is
    \begin{equation}\label{eq:mmse}
        p\left(h_{nt}|\mathbf{Y}\right) = \int p\left(\mathbf{H}, \boldsymbol{\xi}|\mathbf{Y}\right)\, \text{d}\mathbf{\boldsymbol{\xi}}\, \text{d}\mathbf{H}_{\backslash nt}
    \end{equation}

    \noindent where $\bH_{\backslash nt}$ denotes all elements except $h_{nt}$ and the posterior distribution, denoted by $p\left(\bH, \bxi|\bY\right) = \frac{1}{p\left(\bY\right)} p\left(\bY|\bH,\bxi\right) p\left(\bH | \bxi\right) p\left(\bxi\right)$ given by the Bayes' rules
    \vspace{-5pt}
    \begin{align}\label{eq:post_dist}
        p\left(\bH, \bxi|\bY\right)  =& \nicefrac{1}{p\left(\bY\right)} \left[\prod_{l=1}^{L} \prod_{t=1}^{T} p\left(y_{lt} \big| \sum_{n=1}^{N} \phi_{ln}\, h_{nt}\right)\right]\\ \nonumber
        & \times \left[\prod_{n=1}^{N} \prod_{t=1}^{T} P\left(h_{nt}|\xi_{nt}\right)\right] \left[\prod_{n=1}^{N} \prod_{t=1}^{T}P\left(\xi_{nt}\right)\right],
    \end{align}

    \noindent where $P\left(h_{nt}|\xi_{nt}\right)$ is the conditional density for the random variable in~(\ref{eq:H}).

    In order to apply the proposed message scheduling techniques, the first step is to marginalize the problem. As seen in GAMP~\cite{SRangan2011} and HyGAMP~\cite{Rangan2017}, one approach is to employ an approximation of the sum-product loopy belief propagation (BP). For each $t$-th symbol interval, the factor graph (FG) in Fig.~\ref{fig:fg} represents the problem, wherein factor nodes that represents the density functions, prior and likelihood, are depicted as cubes and the variable nodes $\xi_{nt}$ and $h_{nt}$ are seen as spheres. As $\boldsymbol{\Phi}$ is a dense matrix, the FG in Fig.~\ref{fig:fg} is fully connected. Computing the messages in fully connected graphs is tricky as the messages themselves are functions. Thus, a common method is to approximate the messages by prototype functions that resemble Gaussian density functions which can be described by two parameters only. So, message passing reduces to the exchange of the parameters of a function instead of the function itself. Therefore, it is possible to iteratively approximate, for a FG with cycles as in Fig.~\ref{fig:fg}, the marginal posteriors passing messages between different nodes. Thus, we can define the messages from $p\left(y_{lt} \big| \cdot\right)$ to $h_{nt}$ and to the opposite direction as
    \vspace{-5pt}
    \begin{align}\label{eq:mess1}
    \nu^{(i)}_{n\leftarrow lt}\left(h_{nt}\right) \propto& \! \int\! p\left(\!y_{lt} \big| \sum_{k=1}^{N} \phi_{lk}\, h_{kt}\right)\! \times\! \prod_{j\neq n}^N \nu^{(i)}_{j \rightarrow lt}\left(h_{jt}\!\right)\, \text{d}h_{jt} \\[-8pt] \label{eq:mess2}
            \nu^{(i+1)}_{n\rightarrow lt}\left(h_{nt}\right) \propto&  \hspace{5pt} \nu^{(i)}_{n\rightarrow nt}\left(h_{nt}\right)
            \prod_{k\neq t}^T \nu^{(i)}_{n \leftarrow lk}\left(h_{nk}\right)
    \end{align}
    \noindent and, considering $\propto$ as proportional, the messages from $P\left(h_{nt} | \xi_{nt}\right)$ to $h_{nt}$ and to the opposite direction  are
    \vspace{-5pt}
    \begin{flalign}\label{eq:mess3}
        \nu^{(i)}_{n\leftarrow nt}\left(h_{nt}\right) \propto&  \prod_{k=1}^{T} \nu^{(i)}_{n\rightarrow lk}\left(h_{nt}\right),  \\\label{eq:mess4}
        \nu^{(i+1)}_{n\rightarrow nt}\left(h_{nt}\right) \propto &\, \int p\left(h_{nt} | \xi_{nt}\right) \nu^{(i)}_{n\rightarrow nt}\left(\xi_{nt}\right)\, \text{d}\xi_{nt}.
    \end{flalign}

    Thus, the belief distribution that provides an approximation to marginal posterior distribution $p\left(h_{nt} | \bY, \bxi\right)$ is given by
    \begin{equation}\label{eq:Bdist}
        \nu^{(i+1)}_{nt}\left(h_{nt}\right) = \frac{\nu^{(i)}_{n\rightarrow nt} \prod_{s=1}^{T} \nu^{(i)}_{n\leftarrow ls}\left(h_{ns}\right)}{\int \nu^{(i)}_{n\rightarrow nt} \prod_{s=1}^{T} \nu^{(i)}_{n\leftarrow ls}\left(h_{ns}\right)\, \text{d}h_{nt}},
    \end{equation}
    \noindent where, defining $\mathbf{Z} = \bPhi \bH$ and
    \begin{equation}
        \resizebox{0.85\hsize}{!}{
        \hspace{-11pt}$p\left(h \big| \hat{r}_{nt}^{(i)},Q^{r(i)}_{nt};\hat{\rho}^{(i)}_{nt}\right) \triangleq \frac{p\left(h; \hat{\rho}^{(i)}_{nt}\right)\mathcal{C}\mathcal{N}\left(h| \hat{r}_{nt}^{(i)},Q^{r(i)}_{nt}\right)}{\int p\left(h; \hat{\rho}^{(i)}_{nt}\right)\mathcal{C}\mathcal{N}\left(h| \hat{r}_{nt}^{(i)},Q^{r(i)}_{nt}\right)\,\text{d}h}$}
    \end{equation}
    \begin{equation}
        \resizebox{0.85\hsize}{!}{
        \hspace{-11pt}$p\left(z \big| p_{lt}^{(i)},Q^{p(i)}_{lt}\right) \triangleq \frac{p\left(y_{lt}|z^{(i)}_{lt}\right)\mathcal{C}\mathcal{N}\left(z | p_{lt}^{(i)},Q^{p(i)}_{lt}\right)}{\int p\left(y_{lt}|z^{(i)}_{lt}\right)\mathcal{C}\mathcal{N}\left(z | p_{lt}^{(i)},Q^{p(i)}_{lt}\right)\,\text{d}z}$,}
    \end{equation}
    \noindent we can compute $\mathbb{E}\left[\nu^{(i+1)}_{nt}\left(h_{nt}\right)\right] = \hat{h}^{(i+1)}_{nt}$ and $\text{Var}\left[\nu^{(i+1)}_{nt}\left(h_{nt}\right)\right] = Q^{h(i+1)}_{nt}$.

    Specifically, each iteration of Algorithm 1 has three stages. The first one, labelled as ``GAMP approximation'' contains the updates of the GAMP based on expectation propagation (EP) algorithm, which treats the components $h_{nt}$ as independent with the estimated  probability of being active $\hat{\rho}_{nt}$. As well as~\cite{AhnTCom2019, QZou2020}, the EP is incorporated in the process of LBP to the relaxed belief propagation and then to GAMP. At iteration $i$, MSGAMP produces estimates $\hat{\mathbf{h}}^{(i)}$ and $\hat{\mathbf{z}}^{(i)}$ of the vectors $\mathbf{h}$ and $\mathbf{z}$. Several other intermediate vectors, $\hat{\mathbf{p}}^{(i)}$, $\hat{\mathbf{r}}^{(i)}$ and $\hat{\mathbf{s}}^{(i)}$, are also produced. Associated with each of these vectors are matrices like $\mathbf{Q}^{h(i)}$ and $\mathbf{Q}^{z(i)}$ that represent covariances. Thus, in order to reduce the complexity of $O\left(LNT\right)$ to $O\left(NT\right)$, the message in~(\ref{eq:mess1}) is firstly mapped to a Gaussian distribution based on the central limit theorem and Taylor expansions. So, $\nu^{(i)}_{n\leftarrow lt}\left(h_{nt}\right)$ is updated by the Gaussian reproduction property (GRP)~\cite{QZou2020}. Following the same procedure in the messages of~(\ref{eq:mess2}), (\ref{eq:mess3}) and (\ref{eq:mess4}), relaxed BP is obtained by combination of the approximated messages. Since many of these messages slightly differ from each other, in order to fill out those differences, new variables are produced and, ignoring the infinitesimals, GAMP based on EP is obtained.

    The second stage of Algorithm 1, labelled as ``sparsity-rate update'', refers to the ``box'' part of the FG in Fig.~\ref{fig:fg} and updates the estimates of each probability of being active $\hat{\rho}_{ntm}$. In order to use the diversity of the antennas in the BS to refine the activity detection, from this point we include the subscript $m$ into the formulation. Computed using Gaussian approximations of likelihood functions, these estimates are then used to define the message scheduling proposed in this work. The messages in the ``sparsity-rate update'' stage are given by
    }

{
    \vspace{-5pt}
        \begin{align}\label{eq:LBP1}
            \nu^{(i+1)}_{n \leftarrow ntm}\left(\xi_{nt}\right) \propto& \int p\left(h|\xi_{nt}\right) \nu^{(i)}_{n \leftarrow ntm}\left(h\right)\, \text{d}{h}, \\[-5pt] \label{eq:LBP2}
            \nu^{(i)}_{n \rightarrow ntm}\left(\xi_{nt}\right) \propto&\, P\left(\xi_{nt}\right) \prod_{k \neq t}^T \nu^{(i)}_{n \leftarrow nkm}\left(\xi_{nt}\right),
        \end{align}
    \noindent where~(\ref{eq:LBP1}) refers to the message from $P\left(h_{ntm}| \xi_{nt} \right)$ to $\xi_{nt}$ while (\ref{eq:LBP2}) denotes the message in opposite direction and each belief at $\xi_{nt}$ is given by $\nu^{(i)}_n\left(\xi_{nt}\right) \propto P(\xi_{nt})\prod_{t=1}^T \nu^{(i)}_{n\leftarrow ntm}(\xi_{nt})$.

    Defining $\mathcal{X} = R + W$ as a scalar random variable with the same density as $\bH$, the message in (\ref{eq:LBP1}) can be approximated as a likelihood function given by $\nu^{(i)}_{n \leftarrow ntm}\left(\xi_{nt}\right) = \mathcal{C}\mathcal{N}\left(h_{ntm}|\hat{r}_{ntm}^{(i)}, Q^{r(i)}_{ntm} \right)$, where $\hat{r}_{ntm}$ is a component of the AWGN corrupted version of $\mathcal{X}$, $R$, and $Q^r$ is the variance of $\mathcal{X}$. Applying the GRP enables us to define
    \vspace{-5pt}
    \begin{align}\label{eq:LLRnleftnm}
        \text{LLR}_{n\leftarrow ntm}^{(i)}  =& \log \frac{\mathcal{C}\mathcal{N}\left(0 \big| \hat{r}_{ntm}^{(i)}, Q^{r(i)}_{ntm} + \beta_n\right)}{\mathcal{C}\mathcal{N}\left(0 \big| \hat{r}_{ntm}^{(i)}, Q^{r(i)}_{ntm}\right)}.
    \end{align}

    Similarly to~(\ref{eq:LLRnleftnm}), we have $\text{LLR}_{ntm} \triangleq \log \frac{\nu^{(i)}_{n}\left(\xi_{nt} = 1\right)}{\nu^{(i)}_{n}\left(\xi_{nt} = 0\right)}$ and $\text{LLR}_{n\rightarrow ntm}^{(i)} \triangleq \log \frac{\nu^{(i)}_{n \rightarrow ntm}\left(\xi_{nt} = 1\right)}{\nu^{(i)}_{n \rightarrow ntm}\left(\xi_{nt} = 0\right)}$. Substituting~(\ref{eq:LLRnleftnm}) in~(\ref{eq:LBP2}) and in each belief, $\text{LLR}_{n\rightarrow ntm}^{(i)}$ is given by
    \vspace{-5pt}
    \begin{equation}\label{eq:LLRnleftnm2}
        \text{LLR}_{n\rightarrow ntm}^{(i)} = \log \left(\frac{\rho_n}{1-\rho_n}\right) + \sum_{k\neq t}^T \text{LLR}_{n\leftarrow nkm}^{(i)}.
    \end{equation}
   Thereby, the message in~(\ref{eq:mess2}) is described by
    \begin{equation} \label{eq:B_dist}
     \resizebox{0.89\hsize}{!}{$
        \nu^{(i+1)}_{n\rightarrow lt}\left(h_{ntm}\right) = \hat{\rho}^{(i)}_{ntm}\, \mathcal{C}\mathcal{N}\left(h_{ntm}|0,\beta_n\right) + \left(1-\hat{\rho}^{(i)}_{ntm}\right)\delta\left(h_{ntm}\right),    $}
    \end{equation}   \useshortskip
   where
   \begin{equation}\label{eq:est_rho}
        \hat{\rho}^{(i)}_{ntm} \triangleq \nu^{(i+1)}_{n\rightarrow ltm}\left(\xi_{nt} =1\right) = 1 - \nicefrac{1}{1 + \exp\left(\text{LLR}_{n\rightarrow ntm}^{(i)}\right)}.
   \end{equation} \useshortskip
    With the message passing established, the next step is to use the estimates obtained  in~(\ref{eq:est_rho}) in message scheduling.

} 

\input{alg_gdr.tex}

\subsection{Message-scheduling schemes}
    Since it is expected up to $300,000$ devices per cell~\cite{Cisco2020} in future mobile communication systems a technique with low computational cost is fundamental. We develop three different message scheduling criteria that reduce the computational complexity and the number of iterations to reach convergence as compared to HYGAMP.

    MSGAMP determines a group of nodes $\bS^{(i)}$ to update based on two different criterion, the AUD and the RBP. The goal is to update, at every iteration $i$, only the nodes of the group and not all of them, as in HyGAMP. MSGAMP proceeds until $i$ reaches the maximum number of iterations $I$ or $\left(\text{tol}/M < 10^{-4}\right)$, where tol is given by

    \begin{equation}\label{eq:StopCrit}
        \text{tol} = \sum_{m=1}^M \frac{\|\hat{\underline{\bh}}^
        {(i)}_{tm} - \hat{\underline{\bh}}^
        {(i-1)}_{tm}\|}{\|\hat{\underline{\bh}}^
        {(i)}_{tm}\|},
    \end{equation}
    \noindent where $\hat{\underline{\bh}}^{(i)}_{tm}$ is a $|\bS^{(i)}|\times 1$ vector that corresponds to the estimated channel gains between the $|\bS^{(i)}|$ devices and the $m$-th  BS antenna. As this stopping criterion takes into account only the devices in the group, unlike the parallel message update of HyGAMP that, in each iteration, $O(NTM)$ messages must be computed, MSGAMP needs only $O(|\bS^{(i)}|TM)$ . Considering that we have a crowded scenario of MTCDs in future mobile communication systems and the sporadic transmission pattern of each device, the computational cost gain using scheduling schemes is evident, since $|\bS^{(i)}| << N$. With the stopping criterion defined, we present the first message scheduling scheme.

\subsubsection{MSGAMP-AUD}
    The message scheduling based on activity user detection (MSGAMP-AUD) sequentially updates the messages of devices detected as active and repeats the previous values of other devices. The criterion based on AUD uses the estimates of each BS antenna, as $\hat{\rho}_{nt}^{(i)} = \sum_{m=1}^M \hat{\rho}^{(i)}_{ntm}/M$. If $\hat{\rho}_{nt}^{(i)}$  is higher than a threshold, the device is considered as active and is included in the set $\bS^{(i)}$.

    In the first iteration, all messages of all nodes are updated. When $i=2$, we have the first values of $\hat{\boldsymbol{\rho}}_{tm}$, thus enabling the set $\mathbf{S}^{(i)}$. In this iteration, all messages that belong to $\mathbf{S}^{(i)}$, except for $s^{(i)}_1$ will be updated. Then, the index that refers to the messages that had been updated is removed of $\mathbf{S}^{(i)}$ as in
    \begin{equation}
        \mathbf{S}^{(i)} = \left[s^{(i-1)}_2, \dots, s^{(i-1)}_{|\bS^{(i-1)}|}\right].
    \end{equation}
    Therefore, we exclude a group of messages that belong to a specific device to be updated, one at a time. In summary, we reduce the set $\bS^{(i)}$ that is updated in parallel until there is no message to update. When $\bS^{(i)}$ is empty, MSGAMP updates all messages, including the ones that do not belong to the older set, i.e., the new set is $\bS^{(i)} = \left[1, \dots, N\right]$. In the next iteration, a new update of the set $\bS^{(i)}$, using the new $\hat{\rho}_{ntm}$ is performed.

\subsubsection{MSGAMP-RBP}
    In this variation, MSGAMP updates the messages according to an ordering metric called residual belief propagation (RBP). A residual is the norm (defined over the message space) of the difference between the values of a message before and after an update. In our scheme, we define the residual with the beliefs described in~(\ref{eq:Bdist}). Thus, the residual for the belief distribution at $h_{nt}$, is given by
    \begin{equation}\label{eq:Res}
        \text{Res}\left(\nu_{ntm}\left(h_{ntm}\right)\right) = \big|\big|\nu^{(i+1)}_{ntm}\left(h_{ntm}\right) - \nu^{(i)}_{ntm}\left(h_{ntm}\right)\big|\big|.
    \end{equation}

    The intuitive justification of this method is that as the factor graph approach converges, the differences between the messages before and after an update diminish. Therefore, if a message has a large residual, it means that it is located in a part of the graph that has not converged yet. Thus, propagating that message first should speed up the convergence. Using the residual values computed in~(\ref{eq:Res}), we compute the set $\bS^{(i)}$ of messages to be updated in the next iteration. Since the probability of being active of each MTD is typically around $5\%$~\cite{DiRennaAccess2020}, $\bS^{(i)}$ has the $0.05\,N$ nodes with highest residual. The update sequence of MSGAMP-RBP is the same of MSGAMP-AUD, the difference is how both groups are formed.

\subsubsection{MSGAMP-ARBP}

    This dynamic scheduling strategy combines the AUD and RBP criterion. The main idea is use AUD criterion to create $\bS^{(i)}$ and the RBP criterion to compute the updating sequence
    of it. MSGAMP-ARBP updates the messages of one node per iteration, starting with the one with highest residual. After the group being fully updated, MSGAMP-ARBP proceeds as in previous strategies, updating the messages of the nodes that does not belong to $\bS^{(i)}$ and compute a new set. When a stop criterion is met, the activity detection and the channel estimation are given by lines 19 and 5 in Algorithm 1.

\section{Simulation results}
\label{sec:sim}

\begin{figure}[t] 
        \centering
        \includegraphics[scale=0.95]{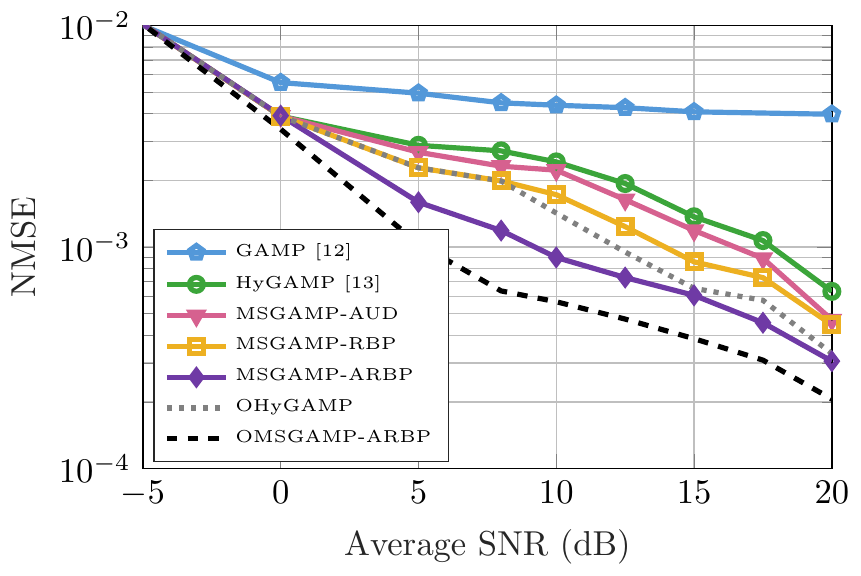}
        \vspace{-10pt}
        \caption{ Normalized mean squared error per frame vs. Average SNR. We considered only the active devices, in the asynchronous scenario with $N=128, M=2$ and $L=32$, after 10 iterations by $10^4$ Monte Carlo trials.}
        \label{fig:NSER}
        \vspace{-10pt}
    \end{figure}

\begin{figure}[t] 
        \centering
        \includegraphics[scale=0.9]{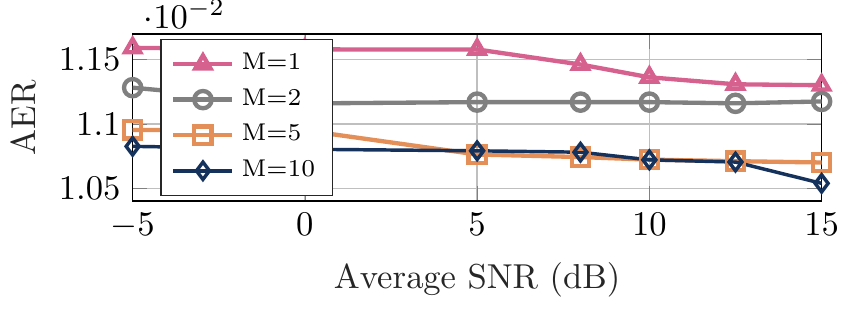}
        \vspace{-10pt}
        \caption{Activity error rate per symbol of MSGAMP-ARBP vs. Average SNR in the asynchronous scenario with $N=128, M=2$ and $L=32$, after 10 iterations. AER is the sum of the missed detections and false alarm rates.}
        \label{fig:AER}
        \vspace{-10pt}
    \end{figure}

\begin{figure}[t]
            \begin{minipage}[b]{1\linewidth}
                \centering
                \centerline{\includegraphics[scale=0.95]{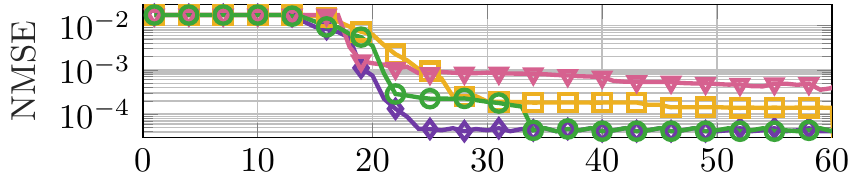}}
                \vspace{-5pt}
                \centerline{\footnotesize (a) SNR = 0 dB.}\medskip
            \end{minipage}

            \begin{minipage}[b]{1\linewidth}
                \centering
                \centerline{\includegraphics[scale=0.95]{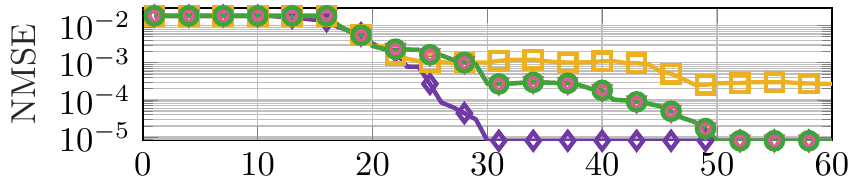}}
                \vspace{-5pt}
                \centerline{\footnotesize (b) SNR = 8 dB.}\medskip
            \end{minipage}

            \begin{minipage}[b]{1\linewidth}
                \centering
                \centerline{\includegraphics[scale=0.95]{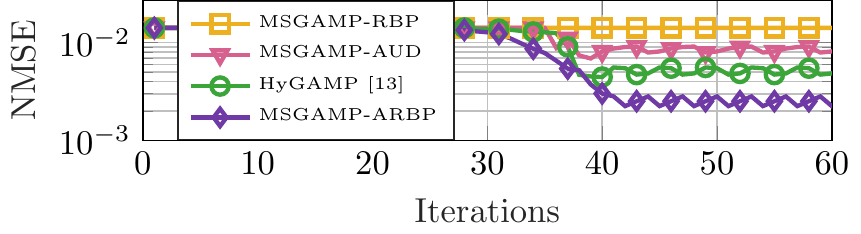}}
                \vspace{-5pt}
                \centerline{\footnotesize (c) SNR = 10 dB.}\medskip
            \end{minipage}
            \vspace{-25pt}
            \caption{ Convergence rate in terms of NMSE per symbol versus iterations. The NMSE considered only the active devices in the asynchronous scenario with $N=128, M=2$ and $L=32$, by $10^4$ Monte Carlo trials.}
            \label{fig:conv}%
            \vspace{-10pt}
        \end{figure}

In order to verify the performance of the proposed MSGAMP schemes,
we simulate an mMTC system with $N=128$ devices, $M=2$ BS antennas,
$L=32$ symbols per frame and $T =3\, L$ as the size of the
observation window. The threshold to detect the activity of devices
considered is 0.9, the average SNR is set to $1/\sigma_w^2$, while
the activity probabilities $p_n$ are drawn uniformly at random in
$\left[0.01,0.05\right]$. The variations of MSGAMP are compared to
the well-known generalized approximate message passing
(GAMP)~\cite{SRangan2011} and the state-of-the-art HyGAMP algorithm
\cite{Rangan2017}. Versions of MSGAMP-ARBP and of HyGAMP with
perfect activity knowledge (OMSGAMP-ARBP and OHyGAMP) are used as
lower bounds. Figs.~\ref{fig:NSER} and~\ref{fig:AER} show results of
NMSE and AER per frame, respectively. In terms of NMSE,
Fig.~\ref{fig:NSER} shows that the message scheduling schemes have a
competitive performance, where MSGAMP-AUD and MSGAMP-RBP slightly
outperform HyGAMP, requiring less computational cost. MSGAMP-ARBP
surpasses not only HyGAMP and the other MSGAMP algorithms but also
OHyGAMP. One can see that the use of the BS antennas in order to
refine the activity detection improved the AER performance of
MSGAMP-ARBP since the AER curves have lower values as $M$ increases.
Fig.~\ref{fig:conv} depicts the convergence rate of MSGAMP-type
techniques and HyGAMP. One can notice that for different values of
SNR, our solutions converge faster and to lower values of NMSE than
HyGAMP. We note that the proposed techniques will be examined with
LDPC codes \cite{memd} in future works.

\section{Conclusion}
\label{sec:conclusion}

In this paper, we have presented a framework for joint activity
detection and channel estimation for mMTC and developed the MSGAMP
algorithm. We have developed three scheduling techniques for MSGAMP
that update the messages based on the AUD and the RBP. The results
indicate that MSGAMP-type techniques outperform other solutions in
terms of NMSE and AER, with fast convergence and low computational
cost.

\bibliographystyle{IEEEbib}
\bibliography{refs}

\end{document}

%% file: alg_gdr.tex
\newcounter{CounterName}
\vspace*{-\baselineskip} 
\begin{table}[t]
	\begin{center} {
		\begin{tabular}{p{8cm}l} \\ \hline 
		{\textbf{Algorithm 1} Message Scheduling GAMP - MSGAMP} \\ \hline
            \textbf{initialize}\\ 
            \stepcounter{CounterName} 
            \hspace{4.5pt}$\theCounterName$:\hspace{5pt} \resizebox{0.95\hsize}{!}{$i=1$, $\hat{{s}}_{ltm}^{(0)} = \hat{{r}}_{ntm}^{(0)} = 0$, $Q^{r(0)}_{ntm} = 1$, $\hat{\rho}^{(0)}_{ntm} = \rho_{n}$, $\bS^{(0)} = \left[1, \dots, N\right]$} \\
            \textbf{repeat}\\
           \hspace{10pt} \% \textit{GAMP approximation}\\
            \stepcounter{CounterName} 
            \hspace{4.5pt}$\theCounterName$:\hspace{5pt} \textbf{for} $\left(n= 1, \dots, |\bS^{(i-1)}|\right) \forall n \in \bS^{(i-1)}$ \\
            \stepcounter{CounterName} 
            \hspace{4.5pt}$\theCounterName$:\hspace{15pt} \textbf{for} $\left(t= 1, \dots, T\right)$ \\
            \stepcounter{CounterName} 
            \hspace{4.5pt}$\theCounterName$:\hspace{25pt} \textbf{for} $\left(m= 1, \dots, M\right)$ \\
            \stepcounter{CounterName} 
            \hspace{4.5pt}$\theCounterName$:\hspace{35pt} $\hat{h}_{ntm}^{(i)} = \mathbb{E}\left[\mathcal{X}_{ntm}\big|\hat{r}_{ntm}^{(i-1)},Q^{r(i-1)}_{ntm};\hat{\rho}^{(i-1)}_{ntm}\right]$ \\[2pt]
            \stepcounter{CounterName} 
            \hspace{4.5pt}$\theCounterName$:\hspace{35pt} ${Q^{h(i)}_{ntm}} = \text{Var}\left[\mathcal{X}_{ntm}\big|\hat{r}^{(i-1)}_{ntm},Q^{r(i-1)}_{ntm};\hat{\rho}^{(i-1)}_{ntm}\right]$ \\[2pt]            
            \stepcounter{CounterName} 
            \hspace{4.5pt}$\theCounterName$:\hspace{35pt} \textbf{for} $\left(l= 1, \dots L \right)$ \\[3pt]
            \stepcounter{CounterName} 
            \hspace{4.5pt}$\theCounterName$:\hspace{45pt} $Q^{p(i)}_{ltm} = \sum_{n=1}^N |\Phi_{ln}|^2 Q^{h(i)}_{ntm}$\\[2pt] 
            \stepcounter{CounterName} 
            \hspace{4.5pt}$\theCounterName$:\hspace{45pt} $p^{(i)}_{ltm} = \sum_{n=1}^N \Phi_{ln}\, \hat{h}_{ntm}^{(i)} - Q^{p(i)}_{ltm} \hat{s}_{ltm}^{(i-1)}$ \\[2pt] 
            \stepcounter{CounterName} 
            \hspace{4.5pt}$\theCounterName$:\hspace{40pt} $\tilde{z}^{(i)}_{ltm} = \left(y_{ltm}\, Q^{p(i)}_{ltm} + \sigma^2_w\, p^{(i)}_{ltm}\right)\big/\left(Q^{p(i)}_{ltm} + \sigma^2_w\right)$\\[2pt]
            \stepcounter{CounterName} 
            \hspace{4.5pt}$\theCounterName$:\hspace{40pt} $Q^{z(i)}_{ltm} = \left(\sigma_w^2\, Q^{p(i)}_{ltm}\right)\big/\left(Q^{p(i)}_{ltm} + \sigma^2_w\right)$ \\[2pt] 
            \stepcounter{CounterName} 
            \hspace{4.5pt}$\theCounterName$:\hspace{40pt} $\hat{s}^{(i)}_{ltm} = \left(\tilde{z}^{(i)}_{ltm}-p^{(i)}_{ltm}\right)\big/Q^{p(i)}_{ltm}$  \\[2pt]
            \stepcounter{CounterName} 
            \hspace{4.5pt}$\theCounterName$:\hspace{40pt} $Q^{s(i)}_{ltm} = Q^{-p(i)}_{ltm} \left(1 - Q^{z(i)}_{ltm}/Q^{p(i)}_{ltm}\right)$  \\[2pt]   
            \stepcounter{CounterName} 
            \hspace{4.5pt}$\theCounterName$:\hspace{30pt} \textbf{end for} \\[2pt] 
            \stepcounter{CounterName} 
            \hspace{4.5pt}$\theCounterName$:\hspace{30pt} $Q^{-r(i)}_{ntm} = \sum_{l=1}^L |\Phi_{ln}|^2 Q^{s(i)}_{ltm}$  \\[2pt]
            \stepcounter{CounterName} 
            \hspace{4.5pt}$\theCounterName$:\hspace{30pt} $r^{(i)}_{ntm} = \hat{h}_{ntm}^{(i)} + Q^{r(i)}_{ntm} \sum_{l=1}^{L} \Phi^\ast_{ln} \hat{s}^{(i)}_{ltm}$ \\[2pt]
            \hspace{12pt} \% \textit{Sparsity-rate update with}~(\ref{eq:LLRnleftnm}), (\ref{eq:LLRnleftnm2}) and~(\ref{eq:est_rho}) \\[1pt]
            \stepcounter{CounterName} 
            \hspace{4.5pt}$\theCounterName$:\hspace{20pt} \textbf{end for} \\[2pt]
            \stepcounter{CounterName} 
            \hspace{4.5pt}$\theCounterName$:\hspace{10pt} \textbf{end for}  \\[2pt]
            \hspace{12pt} \% \textit{Message-scheduling update}\\  
            \stepcounter{CounterName} 
            \hspace{4.5pt}$\theCounterName$:\hspace{5pt} \textit{Refine} $\hat{\rho}_{nt}^{(i)} = \sum_{m=1}^{M}
            \hat{\rho}_{ntm}^{(i)}\big/M$ \\[2pt]
            \stepcounter{CounterName} 
            \hspace{4.5pt}$\theCounterName$:\hspace{5pt} $\bS^{(i)} =$ \textit{update}$\left[\bS^{(i-1)}\right]$ \textit{ with chosen MSGAMP-type technique}  \\[2pt]
            \stepcounter{CounterName} 
            \hspace{4.5pt}$\theCounterName$:\hspace{2pt} \textbf{end for}  \\[2pt]
            \stepcounter{CounterName} 
            \hspace{4.5pt}$\theCounterName$:\hspace{5pt} \textit{Update} $\text{tol}$ \textit{with} (\ref{eq:StopCrit})  and $i = i +1$ \\[2pt]              
            \textbf{until} $\left(i > I \text{ or tol} < 10^{-4} \right)$ \\[1ex]     \hline
		\end{tabular}
		} 
	\end{center}
	\vspace{-20pt}
\end{table} 